\documentclass[twocolumn,secnumarabic,amssymb, nobibnotes, aps, prc, superscriptaddress,nobalancelastpage,nofootinbib]{revtex4-1}

\usepackage{bm}% bold math
\usepackage{amsmath} \usepackage{braket} \usepackage{epsfig}
\usepackage{tensor}
\usepackage[version=4]{mhchem}
\usepackage{titlesec}
\usepackage{ifthen}
\usepackage{sidecap}
\usepackage{listings}
\usepackage[para,online,flushleft]{threeparttablex}
\usepackage{mathtools}
\usepackage{siunitx}
\usepackage[normalem]{ulem}
\usepackage{natbib}
\usepackage{multirow}
\usepackage{enumitem}

\usepackage{xr-hyper}
\usepackage{hyperref}
\hypersetup{breaklinks=true,colorlinks=true,linkcolor=blue,citecolor=blue,filecolor=magenta,urlcolor=cyan}

\usepackage[all]{hypcap}
\usepackage[dvipsnames]{xcolor}

\newcommand{\pk}[1]{ {\color{violet} #1}}

\begin{document}

\title{Impact of ground-state properties and collective excitations on the Skyrme ansatz:\\ 
a Bayesian study
} 

\author{Pietro Klausner}
\email{pietro.klausner@unimi.it}
\affiliation{Dipartimento di Fisica ``Aldo Pontremoli'', Universit\`a degli Studi di Milano, 20133 Milano, Italy}
\affiliation{INFN, Sezione di Milano, 20133 Milano, Italy}
\affiliation{Laboratoire de Physique Corpusculaire L.P.C., CNRS, ENSICAEN, UMR6534, Université de Caen Normandie, CEDEX, 14050 Caen, France}

\author{Gianluca Col\`o}
\email{gianluca.colo@mi.infn.it}
\affiliation{Dipartimento di Fisica ``Aldo Pontremoli'', Universit\`a degli Studi di Milano, 20133 Milano, Italy}
\affiliation{INFN, Sezione di Milano, 20133 Milano, Italy}

\author{Xavier Roca-Maza}
\email{xavier.roca.maza@fqa.ub.es}
\affiliation{Dipartimento di Fisica ``Aldo Pontremoli'', Universit\`a degli Studi di Milano, 20133 Milano, Italy}
\affiliation{INFN, Sezione di Milano, 20133 Milano, Italy}
\affiliation{Departament de F\'isica Qu\`antica i Astrof\'isica, Mart\'i i Franqu\'es, 1, 08028 Barcelona, Spain}
\affiliation{Institut de Ci\`encies del Cosmos, Universitat de Barcelona, Mart\'i i Franqu\'es, 1, 08028 Barcelona, Spain}

\author{Enrico Vigezzi}
\email{enrico.vigezzi@mi.infn.it}
\affiliation{INFN, Sezione di Milano, 20133 Milano, Italy}

\date{\today}

\begin{abstract}
State-of-the-art models based on nuclear Density Functional Theory are successful in the description of nuclei throughout the whole nuclear chart.
Among them, some differences arise regarding their accuracy. For a given nuclear model, this depends on the procedure adopted to determine the parameters, and, at the same time, new experimental findings constantly challenge theory. 
In the present work, we present a Bayesian inference study aimed at assessing the performance of the Skyrme Energy Density Functionals. 
For the sake of simplicity and clarity, we restrict to spherical, double-magic nuclei, giving equal emphasis to ground-state and dynamical properties. 
Our basic constraints are:
i) masses and charge radii, which are known to be very sensitive to the saturation energy and density; 
ii) spin-orbit splittings, which are associated with the spin-orbit parameter(s);
iii) the electric dipole polarizability and parity-violating asymmetry, which are associated with the density dependence of the symmetry energy; 
iv) the excitation energy of the Isoscalar Giant Monopole Resonance, to constrain the nuclear matter incompressibility; 
v) the energy-weighted sum rule of the Isovector Giant Dipole Resonance, to account for the isovector effective mass; 
and vi) the excitation energy of the Isoscalar Quadrupole Resonance, which is related to the isoscalar effective mass.
In this way, we test the Skyrme ansatz in a statistically meaningful way, by determining the posterior distributions of the parameters as well as their correlation and discussing a possible strategy for future developments. 
We are then able to test the Skyrme ansatz in a statistically meaningful way, by determining the posterior distributions of the parameters. In particular, we have found probability distributions in line with published results with the only exceptions of the symmetry energy at saturation $J$ and its slope $L$, whose distributions favor lower values than commonly reported. 
Using our method, we have also been able to discuss the correlations among parameters and observables.
Finally, we discuss in this manuscript a few possible future developments.
\end{abstract}
\maketitle

\section{Introduction}
\label{sec:intro}
One of the most successful approaches to date in describing nuclear properties is Density Functional Theory (DFT) 
\cite{Schunck2019,Colo2020}. The central quantity of such theory is the Energy Density Functional (EDF) which, in principle, gives access to the total energy and the expectation value of any operator. 
However, DFT does not provide the underlying theoretical scheme to build the exact EDF and, currently, only reasonable approximations to the ideal, exact EDF are available. 
In nuclear physics, EDFs have often been built based on effective interactions treated within the mean-field approximation \cite{Bender2003}. 
The most successful effective models are the non-relativistic Skyrme and Gogny EDFs, as well as the relativistic EDFs based either on meson exchange or on point coupling effective interactions \pk{\cite{Vautherin1972, chabanat1997, Decharge1980, Berger1991, Lalazissis, Meng2006, Zhao2010, Zhao2022}}.
Current experimental developments in producing and studying stable and exotic nuclei have allowed, in numerous cases, to determine nuclear properties with an accuracy unreachable for current theory \cite{Roca-Maza2018}. 

Advances in devising new EDFs with better capabilities, accuracy, and predictive power have been reported in the last decade \cite{Schunck2019}. 
However, there exist indications that current EDFs have reached their limits.
Such a claim, in the case of the Skyrme EDFs, has been made when the UNEDF set has been built \cite{Kortelainen2014}.  Within the same Skyrme framework, it has been shown 
that extending the form of the EDF, namely including fourth-order derivative terms on top of the usual second-order ones, does not lead to very significant improvements \cite{Becker2017}. 
A symptom of the obstacles in further improving functionals can be found in the difficulty of reproducing new, different observables. 
A recent example that has attracted considerable interest \cite{Reinhard2021,Essick2021,Reinhard2022,Yuksel2023} is the parity violating asymmetry in ${}^{48}$Ca \cite{CREx} and ${}^{208}$Pb \cite{PREx2}. 
Other examples of current problems can be found in the literature (see, e.g., \cite{Neufcourt2020,Navarro2022}). 

At this stage, and in agreement with other authors, we believe that it is timely to assess the limits of the EDFs, at least in the Skyrme case, with a technique that is as less biased and as statistically meaningful as possible.
A suitable tool for such a goal is to perform a Bayesian inference. The use of Bayesian inference allows estimating the entire probability distributions of the parameter values, while the more traditional $\chi^2$-test only provides point-like estimates for
these parameters, assuming that they are normally distributed. In the recent years, 
Bayesian inference has taken the lead in different fields in physics, and among them also in nuclear physics (see, e.g., \cite{bayesian}). We will discuss our method in detail in Section III. 

The results of the Bayesian inference will depend on the pool of selected data that we wish to reproduce.
Analyzing the sensitivity of the results to this pool is part of our purpose. 
At the same time, we would like to design a methodology based on transparent logic, being well aware that this can be improved in future works. 
Therefore, we try to define a representative set of observables that are known to encode reliable information on the different terms of EDFs. 
Many existing EDFs, although not all, 
have been fitted on masses and radii of a few magic nuclei. We stick to magic nuclei, but we also include some properties of excited states. 
In this way, we hope to be able to test more extensively the different terms of a local functional that depend on the time-even densities and spot possible limitations of the Skyrme ansatz. 

Our first step is to relate, whenever possible, the EDF parameters to properties of the nuclear Equation of State (EoS) at, or around, saturation density $\rho_0$. 
On some of these parameters, empirical information is available \cite{Roca-Maza2018}, and, in addition, they appear to be physically meaningful. 
This strategy has been already used in the literature by different authors (see, e.g., \cite{Stone2007}), and we use a variant of this idea by following the scheme that has been proposed in \cite{Chen2009, Chen2010}. 
This step allows us to use well-founded priors in the Bayesian analysis.

Model parameters related to nuclear surface effects, or the spin-orbit interaction do not have a counterpart in EoS parameters. 
Consequently, those must be treated differently. 
In the literature, there exist slightly different strategies to fix the values of the spin-orbit parameters. 
In the present work, we will adopt the simplest approach, that is, to use two spin-orbit splittings to determine one spin-orbit parameter. 
This restricted selection may introduce a bias.

There is no possibility to easily isolate surface effects on nuclear observables such as masses, radii, or collective excitations, but all these quantities are very sensitive to the surface; in this respect, we will assume and try to test {\em a posteriori} that using physically reliable bulk parameters -- that is, well constrained by reliable priors -- will determine meaningful surface parameters as well. 

We will perform a number of subsequent Bayesian inferences, using each time a larger set of observables. In this way, we will be able to assess more clearly the ability of the employed experimental data, and in particular of those associated with collective excitations,  to constrain the posterior distributions of our selected parameters.

In Sec.~\ref{sec:theo}, we will describe the basic theoretical tools used in this work. 
The Skyrme EDF and the analytic expression for the EoS will be given in Sec.~\ref{sec:skyrme}. 
The connection of the Skyrme EDF parameters with some selected EoS parameters, together with a plausible range of variation for the latter ones, will be detailed in Sec.~\ref{sec:param}. 
The pool of selected observables will be discussed and listed in Sec.~\ref{sec:observables}. 
Since the calculation of nuclear collective states in heavy nuclei is computationally demanding, we will need to resort to a Skyrme model emulator: the main details and a test example will be given in Sec.~\ref{sec:emulator}. 
The basics of the Bayesian inference approach employed in our work will be discussed in Sec.~\ref{sec:bayesian}. 
The results will be found in Sec.~\ref{sec:results}. 
And, finally, our conclusions and perspectives will be drawn in Sec.~\ref{sec:conclusions}.   

\section{Theory}
\label{sec:theo}

In this section, we outline the employed nuclear model, the Skyrme EDF, and its links with the equation of state of nuclear matter.

\subsection{The Skyrme Functional}
\label{sec:skyrme}

The Skyrme EDF can be written as a sum of terms: the kinetic energy of a non-interacting two-fermion system, the terms modeling the effective interaction between nucleons (that includes the spin-orbit part), and the Coulomb energy. 
This translates into the following expression for the energy density $\cal E$, 
\begin{equation}
  \mathcal{E} = \mathcal{E}_{\rm kin} +\mathcal{E}_{\rm int}+\mathcal{E}_{\rm Coul},
\label{eq:edf}
\end{equation}
with
\begin{eqnarray}
  \mathcal{E}_{\rm kin} &=& \frac{\hbar^2\tau_0(r)}{2m^\prime},\\
  \mathcal{E}_{\rm int}&=& \sum_{t=0,1} C_t^\rho\rho_t^2+C_t^{\Delta \rho}\rho_t\Delta\rho_t\nonumber\\&+&C_t^\tau\rho_t\tau_t+\frac{1}{2}C_t^J{\bm J}_t^2+C_t^{\nabla J}\rho_t{\bm\nabla}\cdot{\bm J}_t,\\
  \mathcal{E}_{\rm Coul}&=&2\pi e^2\rho_{\rm ch}(r)\left(\frac{1}{r}\int_0^r\rho_{\rm ch}(r^\prime)r^{\prime 2} dr^\prime\right.\nonumber\\
  &&~~~~~~~~~~~~~~~\left.+\int_r^\infty\rho_{\rm ch}(r^\prime)r^\prime dr^\prime\right)\nonumber\\
  &+& e^2\frac{3}{4}\left(\frac{3}{\pi}\right)^{1/3}\rho_{\rm ch}^{4/3}(r),
  \label{eq:skyrme_functional}
\end{eqnarray}
where the subindex $t=0$ stands for isoscalar and $t=1$ for isovector terms,  and $m^\prime=m A/(A-1)$ includes the one-body part of the center-of-mass correction. 
The $C_t^\rho$ parameter depends on the density as $C_{t0}^\rho+C_{td}^\rho\rho_t^\alpha$, while the other $C$'s are constants (see e.g. Ref.~\cite{Bender2003} for the explicit definitions of the nucleon $\rho_t$, kinetic $\tau_t$ and spin-orbit ${\bm J}_t$ densities). 
In the Coulomb potential written already in spherical symmetry, we approximate the charge density ($\rho_{\rm ch}$) by the proton density ($\rho_p$), and the exchange term is written within the Slater approximation. 

To obtain our results, we will use the code \cite{Colo2021} (see also Ref. \cite{Colo2013}), where the ground-state is obtained by introducing auxiliary single particle orbitals to solve the Schr\"odinger-like equations that can be written after the variational principle is applied to $E=\int d^3{\bm r} \mathcal{E}[\tau_t(r),\rho_t(r), {\bm J}_t(r)]$ (these are the nuclear Kohn-Sham equations). 
Nuclear excited states are calculated based on the same Skyrme EDF through the small amplitude limit of the time-dependent DFT, that is, the so-called Random Phase Approximation or RPA (see, e.g. Ref.~\cite{Nakatsukasa2016}, and Ref.~\cite{Colo2013} for the details of our implementation).
This approach is satisfactory for describing nuclear Giant Resonances, at least as far as their centroid energy or sum rule fraction is concerned. 
We do not aim at reproducing the resonance widths here.  

\subsection{The nuclear Equation of State}
\label{sec:eos}
Due to the relevance of the EoS parameters in our work, we will now briefly introduce the nuclear EoS and give its analytic expression based on the Skyrme EDF ansatz.

In nuclear physics, it is customary to define the EoS as the energy per particle of an ideal infinite system of spin-saturated neutrons and protons at zero temperature (1 MeV $\sim 10^{10}$ K), where the effect of the Coulomb interaction is not taken into account. 
In practice, this means setting to zero the spin-orbit density ${\bm J}_t(r)$ as well as any derivative term of the EDF (uniform system) and the Coulomb contribution (by construction) in Eq.~(\ref{eq:edf}) \cite{chabanat1997}.
Under these assumptions, the energy per particle $E/A \equiv e = \mathcal{E}/\rho_0$ as a function of the total density $\rho_0\equiv\rho=\rho_n+\rho_p$ and isospin asymmetry $\beta\equiv \rho_1/\rho_0 = (\rho_n-\rho_p)/\rho_0$ can be expanded as 
\begin{equation}
e(\rho,\beta)\simeq e(\rho,0) + S_2(\rho)\beta^2 + \mathcal{O}[\beta^4] \ ,       
\end{equation}  
where odd powers of $\beta$ do not appear as a consequence of isospin symmetry.
In the latter equation, the two terms are the energy per particle of symmetric matter $e(\rho,0)$ and the symmetry energy $S_2(\rho)=\frac{1}{2}\frac{\partial^2}{\partial\beta^2} e(\rho,\beta)\vert_{\beta=0}$.
The latter expression up to the second order in $\beta$ has been seen to be quite accurate for densities around and above saturation density \cite{Vidana2009}. 
Hence, it is customary to expand the symmetric matter EoS $e(\rho,0)$ and the symmetry energy $S_2(\rho)$ around the saturation density $\rho_{\rm 0}$, 
\begin{eqnarray}
e(\rho,\beta) &=& E_{\rm 0} + \frac{1}{2}K_{\rm 0}\left(\frac{\rho-\rho_{0}}{3\rho_{0}}\right)^2 + \nonumber \\
              & &  \left[J + L \left(\frac{\rho-\rho_{0}}{3\rho_{0}}\right) + \frac{1}{2}K_{\rm sym}\left(\frac{\rho-\rho_{0}}{3\rho_{0}}\right)^2 \right]\beta^2 \nonumber \\
              & &+ \mathcal{O}[\rho^3,\,\beta^4], 
\end{eqnarray}
and to define the parameters that characterize the density dependence of the EoS and that have some clear physical interpretation.
Those are: i) the saturation density $\rho_{\rm 0}$, obtained from the hydrostatic equilibrium condition in symmetric nuclear matter $P(\rho_{\rm 0})=0$, which essentially determines the size of a nucleus; 
ii) $E_{\rm 0}\equiv e(\rho_{\rm 0},0)$ which has the same physical interpretation as the volume term in the semi-empirical mass formula and, accordingly, plays a role for nuclear binding energies; 
iii) $K_{\rm 0}\equiv 9\rho_{\rm 0}^2\frac{\partial^2}{\partial\rho^2} e(\rho,0)\vert_{\rho=\rho_{\rm 0}}$ which measures the compressibility of symmetric nuclear matter, and has been connected with the excitation energy of the Isoscalar Giant Monopole Resonance \cite{Garg2018}; 
iv) $J\equiv S_2(\rho_{\rm 0})$ which is the symmetry energy at saturation; 
v) $L\equiv 3\rho_{\rm 0}\frac{\partial}{\partial\rho} S_2\vert_{\rho=\rho_{\rm 0}}$ which is proportional to the pressure felt by neutrons in neutron matter at $\rho_{\rm 0}$.
Combinations of $J$ and $L$ are known to be to a good extent related to the properties of isovector nuclear collective excitations, as they explore densities around, and below, saturation \cite{Trippa2008, Roca-Maza2013}.
The $J$ and $L$ parameters are the objects of a lively debate, together with the different types of analysis that can allow to pin down their values; for review papers, the reader can consult Refs. \cite{Baldo2016,Oertel2017,Lattimer2023,Lattimer2023a};
vi) $K_{sym}\equiv 9\rho_{\rm 0}^2 \frac{\partial^2}{\partial\rho^2} S_2(\rho)\vert_{\rho=\rho_{\rm 0}}$, which is the curvature of the symmetry energy. The functional form of the Skyrme interaction is not rich enough to have it as a free parameter, and instead it is entirely determined by $J$ and $L$ \cite{Xu2022}. We include  it for completeness but will not discuss it in what follows.

There are two other useful parameters of the EoS, namely 
vii) the isoscalar $m^*_0$ and 
viii) isovector $m^*_1$ effective masses, ($m^*_t$), which amount to a redefinition of the kinetic energy terms in the EoS, due to potential energy terms that scale as the kinetic energy density $\tau$, namely as $\rho^{2/3}$~\cite{Li2018}. 
It has been found that the Isoscalar Giant Quadrupole Resonance energy is very sensitive to $m_0^*$ \cite{Blaizot1980}, while the energy-weighted sum rule of the Isovector Giant Dipole Resonance (or, more precisely, its enhancement with respect to the classical Thomas-Reiche-Kuhn value) is sensitive to $m_1^*$ \cite{Chabanat1998}.  

\section{Bayesian Inference}
\label{sec:bayesian}
Bayesian inference is a widely used technique for inferring the probability distribution of the parameters of a model given some external information (usually experimental results).

In this section, we will describe our setup: 
the parameters of the  \textit{prior} probability distributions, the experimental observables with their uncertainties, and, finally, the technique we employed to sample the parameters of the  \textit{posterior} probability distribution.

\subsection{Parameters used in the Bayesian inference}
\label{sec:param}
The physical parameters that characterize the nuclear EoS are analytically related to the Skyrme parameters by the following equations  
\begin{widetext}
\begin{eqnarray}
&&P(\rho_0) = 0 = \rho_0^2\frac{\partial e}{\partial \rho}\Big\vert_{\rho_0}=\frac{\hbar^2}{5 m}\left(\frac{3 \pi^2}{2}\right)^{2 / 3} \rho_0^{2 / 3}+C^\rho_0 \rho_0 + \left(\frac{3 \pi^2}{2}\right)^{2 / 3}C^\tau_0 \rho_0^{5 / 3} + C^{\rho}_0(\sigma+1) \rho_0^{\sigma+1}  \\
&&E_0\equiv\frac{E(\rho_0)}{A}=\frac{3 \hbar^2}{10 m}\left(\frac{3 \pi^2}{2}\right)^{2 / 3} \rho_0^{2 / 3}+C^\rho_0 \rho_0+\frac{3}{5} \left(\frac{3 \pi^2}{2}\right)^{2 / 3}C^\tau_0 \rho_0^{5 / 3} + C^{\rho}_0 \rho_0^{\sigma+1} \\
&&K_0\equiv 9\rho_0^2\frac{\partial^2e}{\partial\rho^2}\Big\vert_{\rho_0}=-\frac{3}{5} \frac{\hbar^2}{m}\left(\frac{3 \pi^2}{2}\right)^{2 / 3} \rho_0^{2 / 3}+6 \left(\frac{3 \pi^2}{2}\right)^{2 / 3}C^\tau_0 \rho_0^{5 / 3}+ 9C^{\rho}_0(\sigma+1) \sigma \rho_0^{\sigma+1} \\
&&J\equiv S_2(\rho_0)=\frac{\hbar^2}{6 m}\left(\frac{3 \pi^2}{2}\right)^{2 / 3} \rho^{2 / 3}+C^\rho_1 \rho+\frac{1}{3}\left(\frac{3 \pi^2}{2}\right)^{2 / 3} (C^\tau_0 + 3 C^\tau_1) \rho^{5 / 3}+C^{\rho}_1 \rho_0^{\sigma+1} \\
&&L\equiv 3\rho_0\frac{\partial S_2}{\partial \rho}\Big\vert_{\rho_0}=  \frac{\hbar^2}{3 m}\left(\frac{3 \pi^2}{2}\right)^{2 / 3} \rho_0^{2 / 3}+3C^\rho_1 \rho+\frac{5}{3}\left(\frac{3 \pi^2}{2}\right)^{2 / 3} (C^\tau_0 + 3 C^\tau_1) \rho^{5 / 3} +3(\sigma+1)C^{\rho}_1 \rho_0^{\sigma+1} \\
&&\frac{m_0^*}{m}  =\left(1+ \frac{2m}{\hbar^2} \rho_0 C^\tau_0\right)^{-1} \\
&&\frac{m_1^*}{m}  =\left(1+ \frac{2m}{\hbar^2} \rho_0\left(C^\tau_0-C^\tau_1\right)\right)^{-1}
\end{eqnarray}
\end{widetext}
We will use for the surface parameters $G_0 \equiv-C^{\Delta\rho}_0/2$ for the isoscalar term and $G_1\equiv-C^{\Delta\rho}_1/2$ for the isovector term. 
Also, for simplicity in our notation, we use $W_0$ to indicate the spin-orbit parameter. 
This coincides with the traditional notation of the Skyrme EDF and can be related to the above-introduced parameters as follows: $C_0^{\nabla J}=-3W_0/4$ and $C_1^{\nabla J}=-W_0/4$.    

The quantities defined above are in a one-to-one correspondence with the Skyrme parameters defined above (see Refs.~\cite{Chen2009,Chen2010}).
Working with one or the other is  equivalent; 
however, we have adopted the parametrization in terms of nuclear matter properties because there are definite advantages in working with them (see Sec.~\ref{sec:workflow} below).
We have adopted uniform prior distributions. 
Their boundaries are listed in Table~\ref{tab:prior}. 
Those ranges are based on theoretical analysis of ground and excited state data for the EoS parameters (see Table I in Ref.~\cite{Roca-Maza2018} and references therein) and on the large set of available Skyrme EDFs -- essentially fitted to binding energies and charge radii -- for the surface and spin-orbit parameter. 
We have tried to keep the ranges as large as possible (see further discussions in Sec.~\ref{sec:workflow}).     

\begin{table}
    %\centering
    \caption{
    Intervals for the prior distributions, that have been assumed to be uniform in these intervals.
    }
    \label{tab:prior}
    
    \begin{tabular}{llrr}
    \hline
    & Units & Lower & Upper \\
    && limit & limit \\
    \hline
    $\rho_0$ &[fm$^{-3}$]   &   0.150 &    0.175 \\ 
      $E_0$ &[MeV]       & -16.50 &  -15.50 \\ 
      $K_0$ & [MeV]      & 180.00 &  260.00 \\ 
      $J$ & [MeV]   &  24.00 &    40.00 \\ 
      $L$ &[MeV]         & -20.00 &  120.00\\ 
      $G_0$ &[MeV fm$^5$]&  90.00 &  170.00\\ 
      $G_1$ &[MeV fm$^5$]& -90.00 &   70.00\\ 
      $W_0$ &[MeV fm$^5$] &  60.00 &  190.00\\ 
      $m_0^*/m$&    &   0.70 &    1.10\\ 
      $m_1^*/m$&    &   0.60 &    0.90\\ 
    \hline
    \end{tabular}
    \end{table}

\subsection{Selection of observables and associated errors}
\label{sec:observables}

In Table~\ref{tab:obs}, we show the full list of the experimental observables we selected for the inference. 
In the first part of the Table, we put the ground state properties of several doubly-magic nuclei:
binding energies and charge radii, and two spin-orbit splittings. 
The binding energies have been taken from the AME2020 mass table \cite{AME2020_1, AME2020_2}, while the radii from Ref.~\cite{Angeli2013}.
As for the spin-orbit splittings, we chose the $\nu2p$ splitting of $^{48}$Ca and $\pi2f$ of $^{208}$Pb (from \cite{Zalewski2008}, Tab. III; if more than one value is present, we took the arithmetic mean).
In the second part of the Table, we list the giant isoscalar resonance excitation energies we considered.
We opted for two monopole and one quadrupole resonance: 
for the former, we take $^{208}$Pb, using  data from \cite{PbISGMR} for $E_{\rm GMR}^{\rm IS}$ (constrained energy), and $^{90}$Zr, where the data is from \cite{ZrISGMR} for $E_{\rm GMR}^{\rm IS}$ (still constrained energy);
for the latter, we take $^{208}$Pb $E_{\rm GQR}^{\rm IS}$ (centroid energy), which we took from \cite{PbISGQR}.
Finally, in the third part of the Table, we have listed three isovector properties, which are the dipole polarizability $\alpha_{\rm D}$ of $^{208}$Pb and $^{48}$Ca (from \cite{Tamii2011,Birkhan2017}), the dipole energy-weighted sum (EWSR) rule $m(1)$ of the IVGDR of $^{208}$Pb (from \cite{Tamii2011, Pbm1}) and the parity-violating asymmetry $A_{\rm PV}$ of $^{208}$Pb and $^{48}$Ca (from \cite{PREx2, CREx} and calculated as in Refs.~\cite{Roca-Maza2011-2, Reinhard2021, Reinhard2022}). 
In addition to the experimental values, we provide the total errors adopted for the Bayesian inference.
Binding energies and charge radii are measured to a level of precision far greater than the accuracy of current EDFs.  
Therefore, for those observables, we adopted the errors typical of DFT calculations: 2 MeV for binding energies and 0.05 fm for charge radii. 
We have done the same in the case of  the spin-orbit splittings and the giant resonance energies, to which we have assigned an error of  0.5 MeV.
We used the experimental errors for the isovector properties, taking them from the references reported above.    
\begin{table}
    \centering
    \caption{Observables and initial adopted errors (see text for details). }
    \label{tab:obs}
    \begin{tabular}{cccc}
    \hline
    \multicolumn{4}{c}{Ground-state properties} \\
    \hline
        & $B.E.$ [MeV]   & $R_{\rm ch}$ [fm]   & $\Delta E_{\rm SO}$ [MeV] \\ 
    \hline
    $^{208}$Pb& 1636.4 $\pm$    2.0     &    5.50 $\pm$    0.05  & 2.02 $\pm$    0.50  \\ 
    $^{48}$Ca &  416.0 $\pm$    2.0     &    3.48 $\pm$    0.05  & 1.72 $\pm$    0.50  \\ 
    $^{40}$Ca &  342.1 $\pm$    2.0     &    3.48 $\pm$    0.05  & -                     \\ 
    $^{56}$Ni &  484.0 $\pm$    2.0     & -						 & -					 \\ 
    $^{68}$Ni &  590.4 $\pm$    2.0      & - & -\\ 
    $^{100}$Sn&  825.2 $\pm$    2.0      & - & -\\
    $^{132}$Sn& 1102.8 $\pm$    2.0     &    4.71 $\pm$    0.05  & - \\ 
    $^{90}$Zr & 783.9 $\pm$     2.0     &    4.27 $\pm$    0.05  & - \\
    \end{tabular}\\
    \vspace{0.3cm}
    \begin{tabular}{ccc}
    \hline
    \multicolumn{3}{c}{Isoscalar resonances} \\
    \hline
        & $E_{\rm GMR}^{\rm IS}$ [MeV]  & $E_{\rm GQR}^{\rm IS}$ [MeV] \\
    \hline
    $^{208}$Pb&   13.5 $\pm$    0.5   &  10.9 $\pm$    0.5 \\
    $^{90}$Zr &  17.7 $\pm$ 0.5   & - \\
    \end{tabular}\\
    \vspace{0.3cm}
    \begin{tabular}{cccc}
    \hline
    \multicolumn{4}{c}{Isovector properties} \\
    \hline
        &   $ \alpha_{\rm D}$ [fm$^{3}$]  &   $m$(1) [MeV fm$^2$]  & $A_{\rm PV}$ (ppb) \\
    \hline
    $^{208}$Pb&   19.60 $\pm$    0.60    & 961 $\pm$   22    &      550 $\pm$       18\\
    $^{48}$Ca &    2.07 $\pm$    0.22     & -                 &     2668 $\pm$      113\\
    \hline
    \end{tabular}
 \end{table}

\subsection{Inference and model Emulator}
\label{sec:emulator}
The posterior distribution of the parameters is sampled through the Metropolis-Hastings algorithm, which is a Markov chain Monte Carlo method frequently employed technique for Bayesian inferences.
Given a likelihood function, this implementation of the random walk explores the parameter space, favoring regions of higher likelihood.
In our case, the likelihood is a product of different Gaussian distributions, whose means and widths are the experimental values and the corresponding errors of the observables.
A detailed description of the algorithm can be found in \cite{Hastings1970}. 

The algorithm requires many model evaluations (of the order of $\sim 10^6-10^7$) to reach a satisfactory sampling of the posterior probability distribution of the parameters.
While ground state properties require negligible computational time for this purpose, observables that we must extract from the RPA results are much more computationally demanding, and evaluating them on the fly would be unfeasible.

To overcome this problem, we resorted to the MADAI package, an emulator software based on Gaussian Processes \cite{Rasmussen2006}, built for Bayesian inferences with {\it slow} models.
This tool was developed by the MADAI collaboration (Models and Data Analysis Initiative) \cite{madai}.
The software requires as input a training grid, i.e., a map between representative points in parameter space and the results of the code with those points as input.
Starting from this training grid, the MADAI software builds an emulator of the model, which can be used for the Metropolis algorithm.
For our purposes, we found that $\approx 8500$ points were sufficient for the emulator to behave satisfactorily (see Appendix for more details). 

\subsection{Workflow}
\label{sec:workflow}
\begin{figure}
  \includegraphics[width=0.5\linewidth]{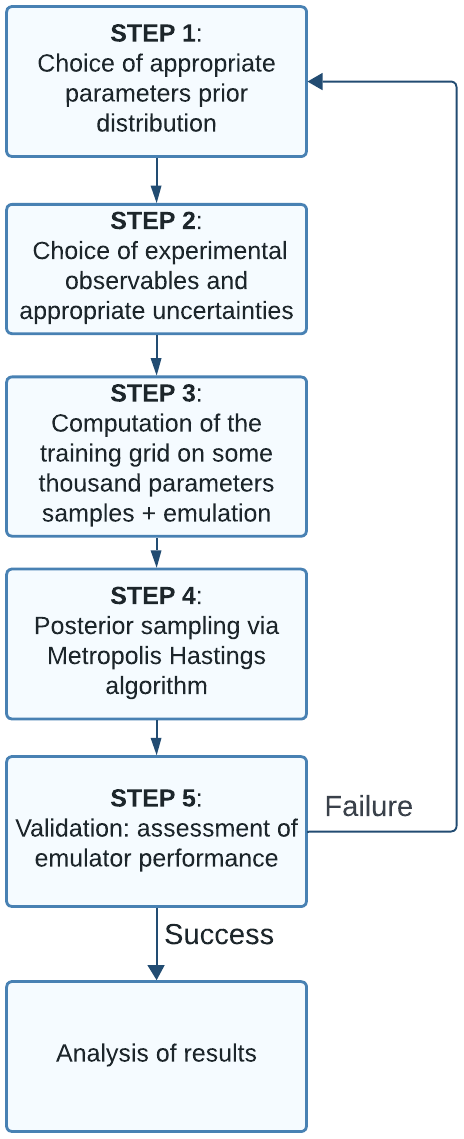}
  \caption{Workflow of the method presented in Sec.~\ref{sec:bayesian} \label{fig-wf}}
\end{figure}

In Fig.~\ref{fig-wf}, we describe the workflow of the method presented here and employed to obtain the results shown in Sec.~\ref{sec:results}. 
The first two steps refer to the creation of the setup: the choice of the observables with their errors (among those listed in Tab.~\ref{tab:obs}) and the prior distribution parameters, which are taken to be uniform inside the intervals listed in Tab.~\ref{tab:prior}.
Then we build the training grid: 
the MADAI software itself proposes a Latin hyper-cube (see appendix of \cite{LatinHypercube}) in parameter space that covers uniformly the space, and we compute the values of the different properties at each point. 

Employing physical parameters described above instead of the parameters of Skyrme interactions is extremely convenient when building the grid since the appropriate range of Skyrme parameters is not evident \textit{a priori}.
Even using seemingly reasonable parameter priors, many combinations of Skyrme parameters belonging to  the training grid will eventually appear to be pathological, i.e., points where our code does not converge. 
Resorting to the nuclear matter parameters drastically decreases the number of these occurrences: a sensible choice of the prior intervals (Tab.~\ref{tab:prior}) excludes most unphysical parameterizations.
Unfortunately, this does not solve entirely the problem, and some points in the training grid remain, for which we cannot compute the experimental observables. These points must be removed, to allow the MADAI software to run.    
In our experience, we have found that we can discard up to $\sim$10\% of the initial grid points without compromising the quality of the emulation.

Once the training grid is ready, the emulator is trained, and the parameter posterior distribution is sampled making use of the Metropolis-Hastings algorithm.
We finally proceed with the validation step, i.e., assessing the performance of the emulator.
We extract a sample of 250 points from the posterior distributions and compare the model results with those obtained by the emulator. 
The validation process is described in detail in Appendix \ref{app:validation}. 
If the validation is not successful, the inference is rejected and the process must start from the beginning, addressing the causes of the poor emulator performance.

\section{Results} 
\label{sec:results}

We turn now to the results.
We use many different observables as input, and we try to constrain many parameters;
therefore, we also try to investigate in some detail how the posterior distributions are affected by specific observables or combinations thereof.

For this reason, we performed seven inferences, all with the same prior distribution (Tab.~\ref{tab:prior}), but by progressively adding the experimental constraints:
\begin{enumerate}
  \item $B.E.\,,\, R_{ch}$: only nuclear masses and radii;
  \item $+\Delta E_{SO}$: masses and radii plus the two spin-orbit splittings (i.e., all the ground state properties);
  \item $+\alpha_D$: the ground state properties plus the nuclear polarizability;
  \item $+GR$: the ground state properties, the nuclear polarizability, the excitation energy of the Isoscalar Monopole and Quadrupole Giant Resonances ($E_{\rm GMR}^{\rm IS}$, $E_{\rm GQR}^{\rm IS}$) and the EWSR of the Isovector Giant Dipole Resonance;
  \end{enumerate} 
  We then add the parity violation asymmetries, first individually, and then together:
  \begin{enumerate}[resume*]
  \item $+A_{PV}$($^{48}$Ca only): the ground state properties, the nuclear polarizability, the Giant Resonances, and the parity-violating asymmetry of $^{48}$Ca;
  \item $+A_{PV}$($^{208}$Pb only): the ground state properties, the nuclear polarizability, the Giant Resonances, and the parity-violating asymmetry of $^{208}$Pb;
  \item $+A_{PV}$ ($^{208}$Pb and $^{48}$Ca): all the observables.
\end{enumerate} 

All those inferences passed the validation step, that is described in detail in appendix \ref{app:validation}.

\subsection{Posterior distribution}
\label{sec:posterior}
\begin{figure*}
  \includegraphics[width=\linewidth]{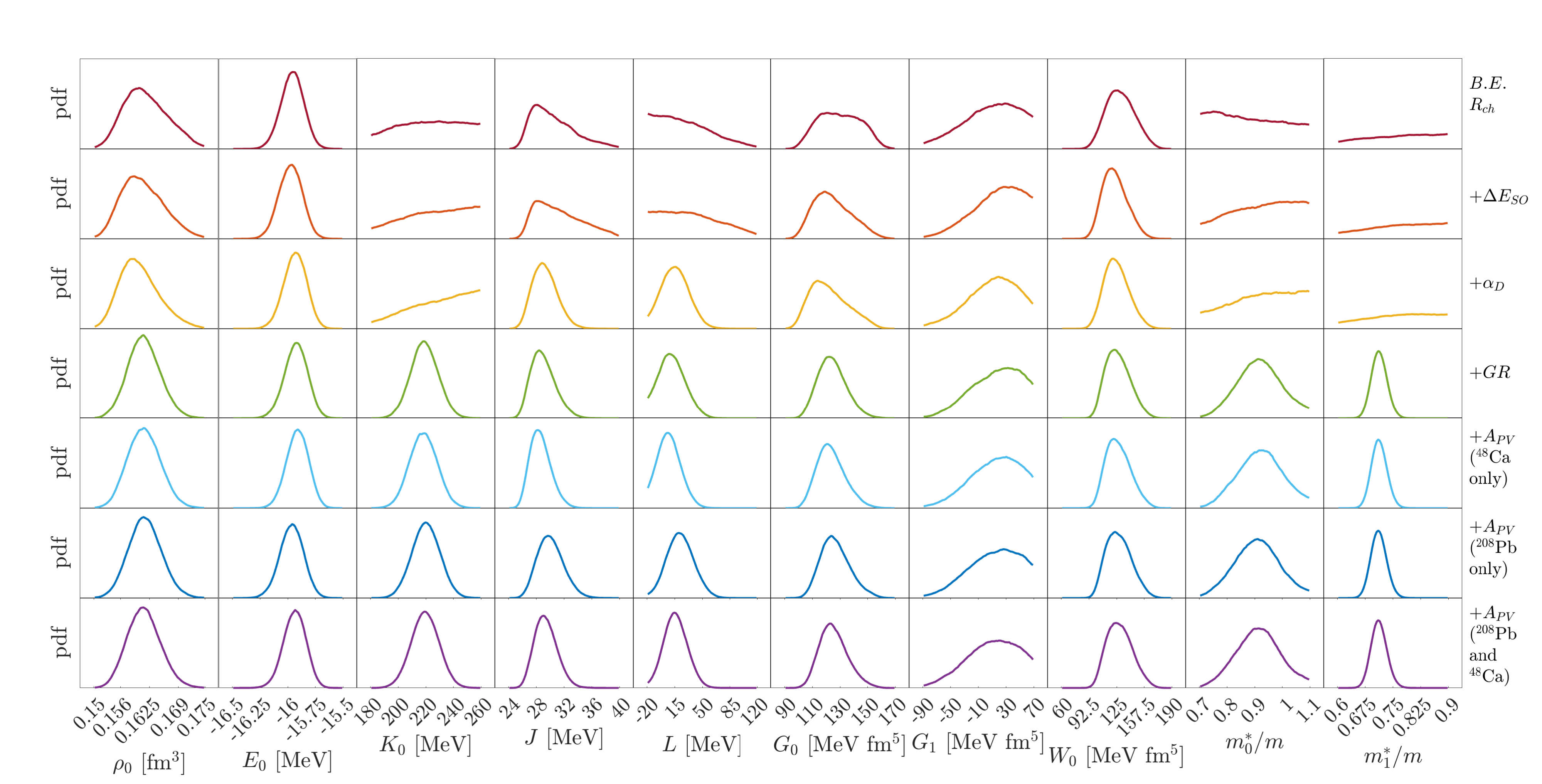}
  \caption{Marginalized posterior distributions of parameters for the seven inferences.}
  \label{fig:posterior_first}
\end{figure*}
\begin{table*}
	\centering
	\footnotesize
	\caption{Means $\mu$ and standard deviations $\sigma$ of the marginalized posterior distributions for all the seven inferences.}
	\label{tab:posterior_mean_std_all}
	\begin{tabular}{llccccccc}
	\hline
	 % & & $B.E.$  & \multirow{2}*{$+ \Delta E_{SO}$} & \multirow{2}*{$+\alpha_D$} & \multirow{2}*{$+GR$} & \multirow{2}*{$+A_{PV}$ $(^{48}Ca)$}   & $+A_{PV}$ $(^{208}Pb)$ & \multirow{2}*{$+A_{PV}$ (Both)}    \\ 
	 % & & $R_{ch}$ &                                    &                              &                       &   & $-A_{PV}$ $(^{48}Ca)$  &                             \\ 
	 & & $B.E.$   & \multirow{2}*{$+ \Delta E_{SO}$}   & \multirow{2}*{$+\alpha_D$}   & \multirow{2}*{$+GR$}  & $+A_{PV}$         & $+A_{PV}$          & $+A_{PV}$ ($^{208}$Pb   \\ 
	 & & $R_{ch}$ &                                    &                              &                       & ($^{48}$Ca only) & ($^{208}$Pb only)  &  and $^{48}$Ca)   \\ 
	\hline
	\multirow{2}*{$\rho_{0}$} & $\mu$ & $   0.162$ & $   0.161$ & $   0.160$ & $   0.161$ & $   0.161$ & $   0.162$ & $   0.161$ \\ 
	  & $\sigma$ & $   0.005$ & $   0.005$ & $   0.004$ & $   0.004$ & $   0.004$ & $   0.004$ & $   0.004$ \\ 
	  \cline{2-9} 
	\multirow{2}*{$E_{0}$} & $\mu$ & $  -15.96$ & $  -15.97$ & $  -15.93$ & $  -15.93$ & $  -15.91$ & $  -15.96$ & $  -15.94$ \\ 
	  & $\sigma$ & $    0.10$ & $    0.11$ & $    0.10$ & $    0.11$ & $    0.10$ & $    0.11$ & $    0.10$ \\ 
	  \cline{2-9} 
	\multirow{2}*{$K_{0}$} & $\mu$ & $     223$ & $     226$ & $     229$ & $     219$ & $     218$ & $     220$ & $     219$ \\ 
	  & $\sigma$ & $      22$ & $      22$ & $      21$ & $      10$ & $      10$ & $      10$ & $      10$ \\ 
	  \cline{2-9} 
	\multirow{2}*{$J$} & $\mu$ & $    30.5$ & $    31.3$ & $    29.2$ & $    29.0$ & $    28.6$ & $    30.1$ & $    29.4$ \\ 
	  & $\sigma$ & $     3.1$ & $     3.4$ & $     1.8$ & $     1.8$ & $     1.5$ & $     1.9$ & $     1.6$ \\ 
	  \cline{2-9} 
	\multirow{2}*{$L$} & $\mu$ & $    28.4$ & $    36.1$ & $    15.8$ & $    11.8$ & $     8.2$ & $    22.2$ & $    16.1$ \\ 
	  & $\sigma$ & $    33.1$ & $    35.8$ & $    17.2$ & $    16.5$ & $    14.4$ & $    16.9$ & $    14.7$ \\ 
	  \cline{2-9} 
	\multirow{2}*{$G_0$} & $\mu$ & $     130$ & $     125$ & $     122$ & $     124$ & $     124$ & $     127$ & $     125$ \\ 
	  & $\sigma$ & $      15$ & $      14$ & $      14$ & $      10$ & $      10$ & $      11$ & $      10$ \\ 
	  \cline{2-9} 
	\multirow{2}*{$G_1$} & $\mu$ & $       8$ & $      16$ & $      10$ & $      14$ & $      13$ & $      12$ & $       9$ \\ 
	  & $\sigma$ & $      38$ & $      35$ & $      35$ & $      35$ & $      34$ & $      36$ & $      36$ \\ 
	  \cline{2-9} 
	\multirow{2}*{$W_0$} & $\mu$ & $     130$ & $     123$ & $     125$ & $     127$ & $     127$ & $     128$ & $     129$ \\ 
	  & $\sigma$ & $      17$ & $      15$ & $      15$ & $      14$ & $      14$ & $      15$ & $      15$ \\ 
	  \cline{2-9} 
	\multirow{2}*{$m^*_0/m$} & $\mu$ & $    0.88$ & $    0.93$ & $    0.93$ & $    0.92$ & $    0.92$ & $    0.91$ & $    0.91$ \\ 
	  & $\sigma$ & $    0.12$ & $    0.11$ & $    0.11$ & $    0.08$ & $    0.08$ & $    0.08$ & $    0.08$ \\ 
	  \cline{2-9} 
	\multirow{2}*{$m^*_1/m$} & $\mu$ & $    0.76$ & $    0.77$ & $    0.77$ & $    0.71$ & $    0.71$ & $    0.71$ & $    0.71$ \\ 
	  & $\sigma$ & $    0.08$ & $    0.08$ & $    0.08$ & $    0.02$ & $    0.02$ & $    0.02$ & $    0.02$ \\ 
	  \cline{2-9} 
	\hline
	\normalsize	\end{tabular}\\ 
\end{table*}

We will now discuss the posterior distribution of the parameters and study how it evolves by adding more and more observables.
In Fig.~\ref{fig:posterior_first}, we show the posterior distribution marginalized over all but one of our ten parameters.
Each of the seven rows of the figure corresponds to one of the seven inferences listed above, going from top to bottom.
In each of the ten columns, we display the sampled probability distribution function (pdf) associated with the entire prior interval of the corresponding parameter. 
For the sake of illustration, the means and standard deviations of each distribution are reported in Tab.~\ref{tab:posterior_mean_std_all}. 
However, they are of limited or no significance when the distributions tend to be flat over the prior interval. 

In the first row, we can observe that the masses and radii constrain effectively the energy at saturation $E_0$, the saturation density $\rho_0$ and the spin-orbit parameter $W_0$.
Adding the spin-orbit constraint in the second row lowers a little the mean value and the width of the $W_0$ distribution, and has a small influence on the other parameters. 
It is worth noting that our choice of spin-orbit splittings seems to slightly favor values of $m^*_0$ in the upper part of the prior interval. 
The posterior distributions are instead almost flat for $K_0$ and the isoscalar and isovector effective masses.

The $G_1$ parameter distribution, which remains fairly identical in all the seven inferences, peaks within the boundaries of the prior interval, but does not have enough space to develop its tail.
We tried to enlarge the prior interval, but this created several pathological points in the emulator training grid, well above the empirical 10\% ``safety limit'' that we had set (see the discussion above in \ref{sec:workflow}).
We checked that the vast majority of these points had $G_1 > 70$ MeV fm$^{5}$.
The isoscalar surface parameter $G_0$ displays a  broad distribution  that becomes sharper once more observables are added.

The distribution of the $J$ and $L$ parameters appear to be weakly constrained by these two first inferences. 
The $J$ distribution peaks slightly below 28 MeV, and has a mean value of about 31 MeV and a standard deviation  of 3 MeV, displaying a long tail which explores the full prior interval up to 40 MeV. The $L$ distribution is rather flat.
On  the other hand,  we do observe the  strong correlation between $J$ and $L$ which is well documented in the literature (see for example \cite{Lattimer2013}).
In fact, in the second inference the two parameters have a correlation coefficient of 0.96, in line with other works. We will be back to this point at the end of this subsection.

Our distributions can be compared  with other investigations  in which  only  ground state constraints  have been  considered.
In \cite{Kortelainen2010} a set  of binding energies and charge radii from 72 nuclei, both closed-shell and open-shell, and both spherical and deformed was used in the analysis.   The optimal value  found  for $J$ in the case of the   the UNEDF0 functional  (30.54 $\pm$ 3.06 MeV)  is compatible with ours, while $L$ was poorly constrained.
In \cite{Moller2012}, the full AME2003 mass table was analyzed with the finite-range droplet model (FRDM), obtaining an error of
$\sigma= 0.57$ MeV on the masses and the  optimal values $J = 32.5 \pm 0.5 $ MeV and $L = 70 \pm 15 $ MeV.   These values are compatible with those obtained in our $\Delta E_{SO}$ inference, but are determined with a much smaller error, showing that 
including  more binding energies and charge radii may be helpful in future Bayesian analysis.

Our picture changes drastically when we add the nuclear polarizability $\alpha_D$ in the pool of observables: 
now $J$ and $L$ are well constrained. In the literature, the correlation between $L$, $J$ and $\alpha_D$ is amply documented.
For example, in \cite{Roca-Maza2013_2} the authors have found a strong correlation between the slope $L$ and the product $\alpha_DJ$.
Furthermore, in \cite{Roca-Maza2015}, the authors expanded their work by adding the experimental data of $^{68}$Ni and $^{120}$Sn $\alpha_D$ as well, and wrote explicitly a linear relation between $L$ and $J$ using the experimental values.
They also remarked that the functionals able to reproduce the experimental values of $\alpha_DJ$ in $^{68}$Ni, $^{120}$Sn and $^{208}$Pb within 1 $\sigma$ have $J\in[30,35]$ MeV and $L\in[20,66]$ MeV.

The observables introduced with the $GR$ inference (fourth row of Fig. \ref{fig:posterior_first}, green lines) have multiple effects: the excitation energies of the GMR and GQR  constrain $K_0$ and  $m^*_0/m$ respectively, while the $m(1)$ value for the IVGDR constrains $m^*_1/m$.

Let us analyze the impact of the monopole constraint.
We find that the mean value of $K_0$ is approximately $220$ MeV, with a standard deviation of $10$ MeV. These values are not 
affected by the addition of further observables (Tab. \ref{tab:posterior_mean_std_all}).
Our result is compatible with  previous analyses, which  deduced a value $K_0=240\pm20$ MeV \cite{Shlomo2006} by comparing with the ISGMR experimental results, taking into account 
the fact that Skyrme EDFs may have different density dependences (cf. also \cite{Colo2004}), and also considering relativistic EDFs.  
While such analyses have been mainly based on $^{208}$Pb, other nuclei may point to slightly lower values of $K_0$ \cite{Garg2018}.
Other studies  \cite{Khan2012} pointed out that medium-heavy nuclei have a mean density that is lower than saturation (typically, around $\rho \approx 0.1$ fm$^{-3}$).
By analyzing the giant monopole resonance data from this perspective, they predicted a less stringent interval of $K_0=230\pm40$ MeV.

The isoscalar effective mass $m^*_0/m$ is constrained once we include the excitation energy of the $E_{GQR}^{IS}$ in $^{208}$Pb, leading to relatively high values ($m^*_0/m \approx $ 0.9). 
This confirms previous findings, starting from the pioneering work of Ref. \cite{Blaizot1980}. 
More recently  \cite{Roca-Maza2013}, it was shown that models with lower values of the effective mass tend to predict too high excitation energies for the collective mode.
The EWSR of the IVGDR instead constrains the isovector effective mass $m^*_1/m$ because both quantities are connected with the so-called isovector enhancement factor \cite{chabanat1997,Roca-Maza2018}.

The dipole polarizability and the energy of the  ISGMR in $^{208}$Pb were also used in  \cite{Yuskel2019} to fit a relativistic energy density functional, in addition to the ground state properties of many nuclei, including open-shell nuclei with their pairing correlations. The resulting DD-PCX functional yields $J= 31.13 \pm 0.32$ MeV and $L= 46.32 \pm 1.68$ MeV.
As already mentioned, having a larger pool of nuclei is certainly an asset; at the same time, we cannot directly compare the latter errors with ours, as the fit was performed by a $\chi^2$ minimization. 

\begin{figure}
\includegraphics[width=\linewidth]{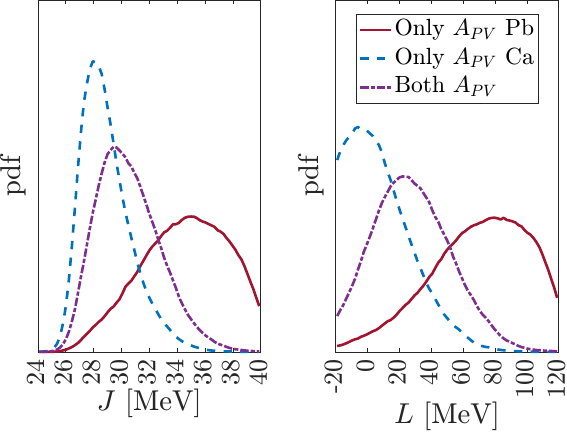}
  \caption{Effect of the $A_{PV}$ on the posterior distributions of $J$ and $L$, if $\alpha_D$ is not in the observables pool.
           We can observe the tension between the  experimental results in $^{48}$Ca and $^{208}$Pb.}
  \label{fig:J_L_only_apvs_both}
\end{figure}

We included the parity-violating asymmetries in the last three inferences.
As we can observe in Fig.~\ref{fig:posterior_first}, adding the $A_{PV}(^{48}$Ca) (light blue lines) or the $A_{PV}(^{208}$Pb) (blue lines) has opposite effects on the $J,\,L$ distributions, shifting them towards slightly lower or slightly higher values, respectively. This tendency becomes much more pronounced if the polarizability is excluded from the pool of observables. 
This can be clearly seen in Fig.  \ref{fig:J_L_only_apvs_both}, where we show the posterior distributions associated with $J$ and $L$, obtained by excluding $\alpha_D$ from the inference. 
Including only the $^{208}$Pb $A_{PV}$, the $J$-distribution peaks around 36 MeV while the $L$-distribution peaks around 85 MeV. 
This is consistent with previous works showing that a high $L$ value is needed to explain the PREX-II results \cite{PREx2, Reinhard2021}.
On the other hand, including only the $^{48}$Ca $A_{PV}$ the $J$-distribution peaks around 28 MeV while the $L$-distribution peaks around 0 MeV. 
  
This is the manifestation of the known tension between the two measurements in $^{48}$Ca and $^{208}$Pb \cite{Reinhard2022,Yuksel2023}.
It must be pointed out any way that the distributions are quite broad, and they overlap over a rather extended region of the parameter space. 
Including both $A_{PV}$'s one obtains distributions intermediate between the two extreme cases that we have discussed, with $J$ and $L$ peaked around 30 MeV and 20 MeV respectively.  
These values are not far from those obtained in the final $+A_{PV}$ inference including all our observables, shown in Fig.~\ref{fig:posterior_first}, which however displays smaller widths, due to the polarizability constraint.  

In a recent work \cite{Zhang2023}, the authors performed a Bayesian analysis similar in spirit to ours, also based on Skyrme interactions but including a less diversified set of observables. 
Their constraints include ground state properties of doubly magic nuclei, and the excitation energy of the $^{208}$Pb monopole giant resonance; they also introduced the PREX-II and CREX results in the form of the deduced weak form factors, and not directly with the experimental observable $A_{PV}$. 
On the other hand, $\alpha_D$, the IVGDR energy weighted sum rule, and the excitation energy of the ISGQR were not included in the inference.

Their resulting posterior distributions for $J$ and $L$ are similar to those we obtain neglecting $\alpha_D$, shown above in Fig. \ref{fig:J_L_only_apvs_both}. 
They find $J=29.1\pm^{2.1}_{1.8}$ MeV and $L=17.1\pm^{23.8}_{22.3}$ MeV  (68.3\% credible intervals).  
Their values for $K_0$ lie in the interval $225\pm^{2.9}_{2.8}$ MeV (68.3\% credible interval), which is compatible with our result. 
They also produced a representative Skyrme interaction, SkREx, whose parameters lie well within our posterior distributions, except for $G_1 = 55$ MeV fm$^5$, a value which is slightly disfavored by our findings.
This interaction predicts values of $\alpha_D$ lying within $1\sigma$ of the experimental results for both $^{48}$Ca and $^{208}$Pb.

Another Bayesian study was performed in \cite{Xu2022}, based on previous exploratory studies \cite{Xu2020, Xu2021_CPL, Xu2021_PRC}, both with the traditional Skyrme interaction and with KIDS functionals \cite{Gil2017, Gil2017_ActaPolonica, Papakonstantinou2018}.
The results of this study in the isovector sector, and especially the values  of $J$ and $L$, differ from our findings.
However, their fitting protocol is different from ours:
the authors fix some parameters ($E_0$, $\rho_0$, $G_0$, $G_1$, $W_0$ and $m_0^*/m$);
they perform inferences with observables from either $^{208}$Pb or $^{120}$Sn; 
they do not use a gaussian likelihood for B.E. and charge radii of the aforementioned nuclei, but rather they accept any parameters combination that predicts them within 3\% of the experimental value;
they use the data of the PREX-II experiment in terms of inferred neutron skin and not of the experimental observable $A_{PV}$;
finally, they use a different dataset (i.e., photoabsorbtion data) for the $^{208}$Pb IVGDR EWSR.
Understanding the impact of these differences is not evident \textit{a priori}, and it is left to future investigations.

In \cite{Sun2024}, the authors compiled the parameters and nuclear matter properties of 255 published Skyrme interactions.
Thus, it is interesting to compare our parameters marginalized posterior distributions with those of published parametrizations, even though the statistical meaning of the latter is not as clear as ours.
We start with the saturation density $\rho_0$.
The distributions are quite similar, even though most Skyrme have $\rho_0 \approx 0.16$ fm$^{-3}$, whereas ours is slightly shifted to higher values.
On the other hand, we find that $E_0$ is in line with what is published, with most Skyrme EDFs having values in the interval $(-16.3,\;-15.5)$ MeV.
Instead, $K_0$ is lower than commonly given values, peaking around 220 MeV and having tails below 200 MeV, while most Skyrme interactions have $K_0$ values around $235-240$ MeV.
Similarly, but more markedly, we find lower values than previously reported for $J$ and especially for $L$.
We find that $J$ peaks at around 29 MeV, and values over 32 MeV are heavily disfavored, while most Skyrme have $J$ between 31 and 33 MeV.
For $L$, we find a peak at around 15 MeV, while values over 45 MeV are highly unlikely; instead, published Skyrme show a much wider interval, that ranges from 0 MeV to more than 100 MeV, with those around 50 MeV particularly frequent.
As for the surface parameters, we find published $G_0$ values ranging from 0 to 200 MeV fm$^5$, even though the most frequent are those where our distribution peaks.  
Instead, our $G_1$ distribution and the published one are very similar, which is quite reassuring given the fact that we should - but could not - expand its prior interval.
For both effective masses $m^*_t/m$ we find a much tighter spread in values. 
Both of the published ones span from 0.5 to slightly more than 1.
Finally, the authors of Ref. \cite{Sun2024} did not report the values of the spin-orbit parameter.
On the other hand, given our experience, our results, which are centered around 130 MeV fm$^5$ are in line with the most common parametrizations.

\begin{figure*}
  \includegraphics[width=\linewidth]{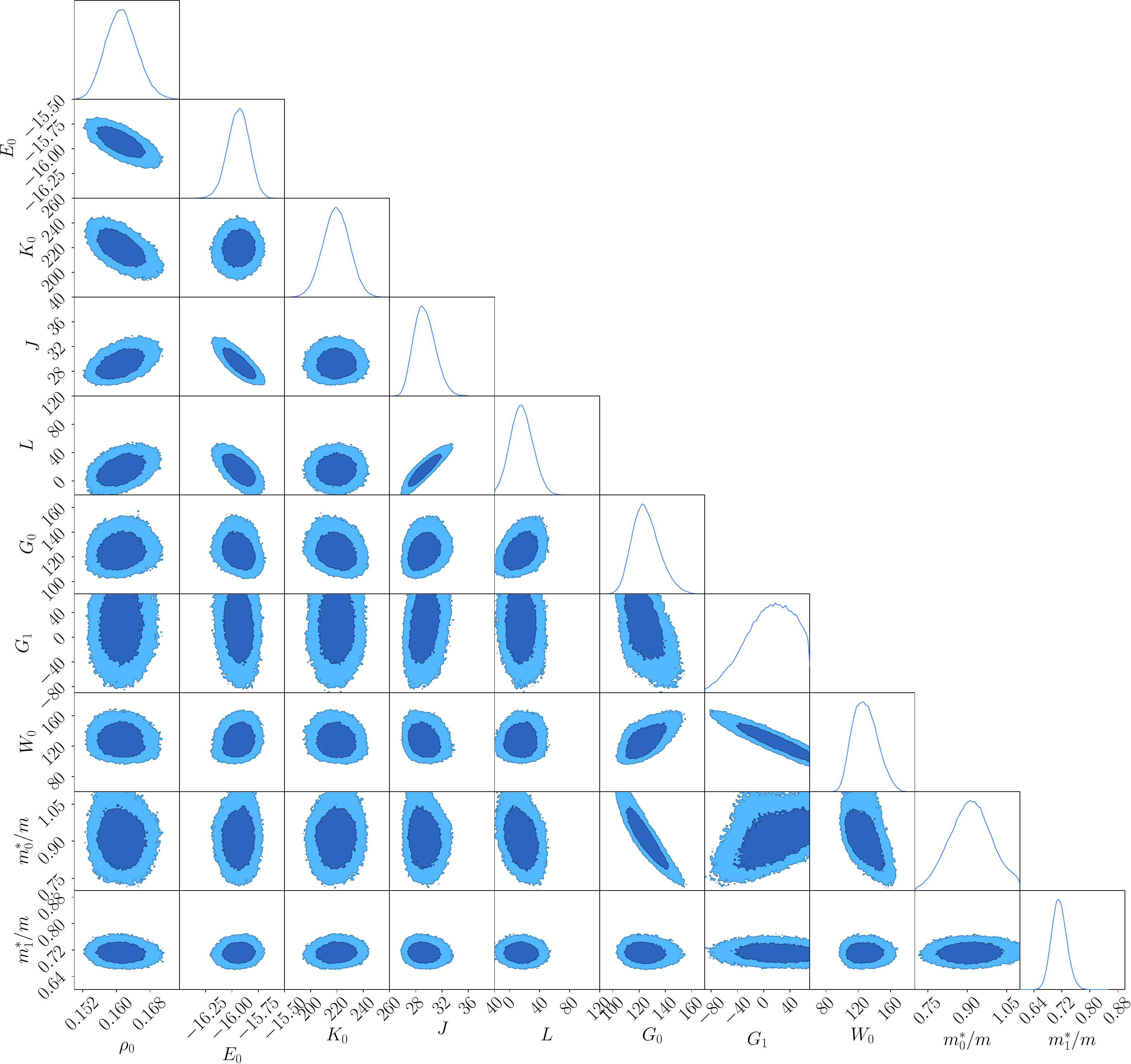}
  \caption{Corner plot when using all observables.}
  \label{fig:corner_plot}
\end{figure*}

We conclude this Section by showing the corner plot associated with our
final $+A_{PV}$ inference in Fig.~\ref{fig:corner_plot}.
The single parameter marginalized posterior distributions are shown on the diagonal, while the other panels contain the combined marginalized distributions. 
This plot is useful to identify the strongest correlations among the observables. 
In particular, one can notice the strong correlation between $J$ and $L$ already discussed above. 
The correlation between $m^*_0/m$ and $G_0$ results from the fact that they are different combinations of the same Skyrme parameters. Other strong correlations are observed between $J,L$ and $E_0$, and between $W_0$ and $G_1$.

\subsection{Observables from the  posterior distribution}
\label{sec:observables_posterior}
It is interesting to investigate how well the  Skyrme parametrizations follow the posterior distribution to reproduce the observables used for the fit.
For this analysis, we extracted 100000  parameter samples and emulated the model results.
From this distribution, we computed the arithmetic mean and standard deviation of observables along the samples. The results are collected in Tab.~\ref{tab:full_run_new_results}.
\begin{table}
	\centering
	\caption{Mean and standard deviation of the observables posterior distributions.}
	\label{tab:full_run_new_results}
	\begin{tabular}{cccc}
	\hline
	\multicolumn{4}{c}{Ground-state properties} \\ 
	\hline
	   & $B.E.$ [MeV] & $R_{\rm ch}$ [fm]& $\Delta E_{\rm SO}$ [MeV]  \\ 
	\hline
	$^{208}$Pb & $    1636 \pm      1.8$ & $    5.49 \pm     0.03$ & $    2.34 \pm     0.16$  \\ 
	$^{48}$Ca & $     417 \pm      1.2$ & $    3.51 \pm     0.02$ & $    1.92 \pm     0.20$  \\ 
	$^{40}$Ca & $     342 \pm      1.6$ & $    3.50 \pm     0.02$ & -  \\ 
	$^{56}$Ni & $     482 \pm      1.4$ & - & - \\ 
	$^{68}$Ni & $     590 \pm      1.0$ & - & - \\ 
	$^{100}$Sn & $     826 \pm      1.6$ & - & - \\ 
	$^{132}$Sn & $    1103 \pm      1.7$ & $    4.71 \pm     0.03$ & -  \\ 
	$^{90}$Zr & $     784 \pm      1.3$ & $    4.27 \pm     0.02$ & -  \\ 
	\end{tabular}\\ 
	\vspace{0.3cm}
	\begin{tabular}{ccc}
	\hline
	\multicolumn{3}{c}{Isoscalar resonances} \\ 
	\hline
	   & $E_{\rm GMR}^{\rm IS}$ [MeV] & $E_{\rm GQR}^{\rm IS}$ [MeV] \\ 
	\hline
	$^{208}$Pb & $    13.5 \pm      0.3$ & $    10.8 \pm      0.4$  \\ 
	$^{90}$Zr & $    17.8 \pm      0.4$ & - \\ 
	\end{tabular}\\ 
	\vspace{0.3cm}
	\begin{tabular}{cccc}
	\hline
	\multicolumn{4}{c}{Isovector properties} \\ 
	\hline
	   & $\alpha_D$ [fm$^3$] & $m(1)$ [MeV fm$^2$] & $A_{PV}$ [p.p.b.] \\ 
	\hline
	$^{208}$Pb & $    19.5 \pm       0.5$ & $     958 \pm        22$ & $     589 \pm         5$ \\ 
	$^{48}$Ca & $    2.30 \pm      0.08$ & - & $    2591 \pm        54$ \\ 
	\hline
	\end{tabular}\\ 
\end{table}

We find that almost all our results lie within 1$\sigma_c$ from the experimental data, where $\sigma_c = \sqrt{ \sigma_{inf}^2 + \sigma_{exp}^2}$, and $\sigma_{inf}$ is the standard deviation of the resulting posterior distribution while $\sigma_{exp}$ is the experimental error (even for those observables to which we assigned a theoretical error for the inference).

The only exceptions are the binding energy of the $^{56}$Ni and the spin-orbit splitting of $^{208}$Pb, which are between 1 and 2$\sigma_c$, and the $^{208}$Pb $A_{PV}$, that lies at slightly more than 2$\sigma_c$ (2.08).
This is in keeping with our previous discussion (see Fig.  \ref{fig:J_L_only_apvs_both})  about the tension between  $^{208}$Pb $A_{PV}$ and $^{48}$Ca $A_{PV}$ and the dominant effect of $\alpha_D$ which leads to low values for $J$ and $L$.

\begin{figure*}
  \includegraphics[width=\linewidth]{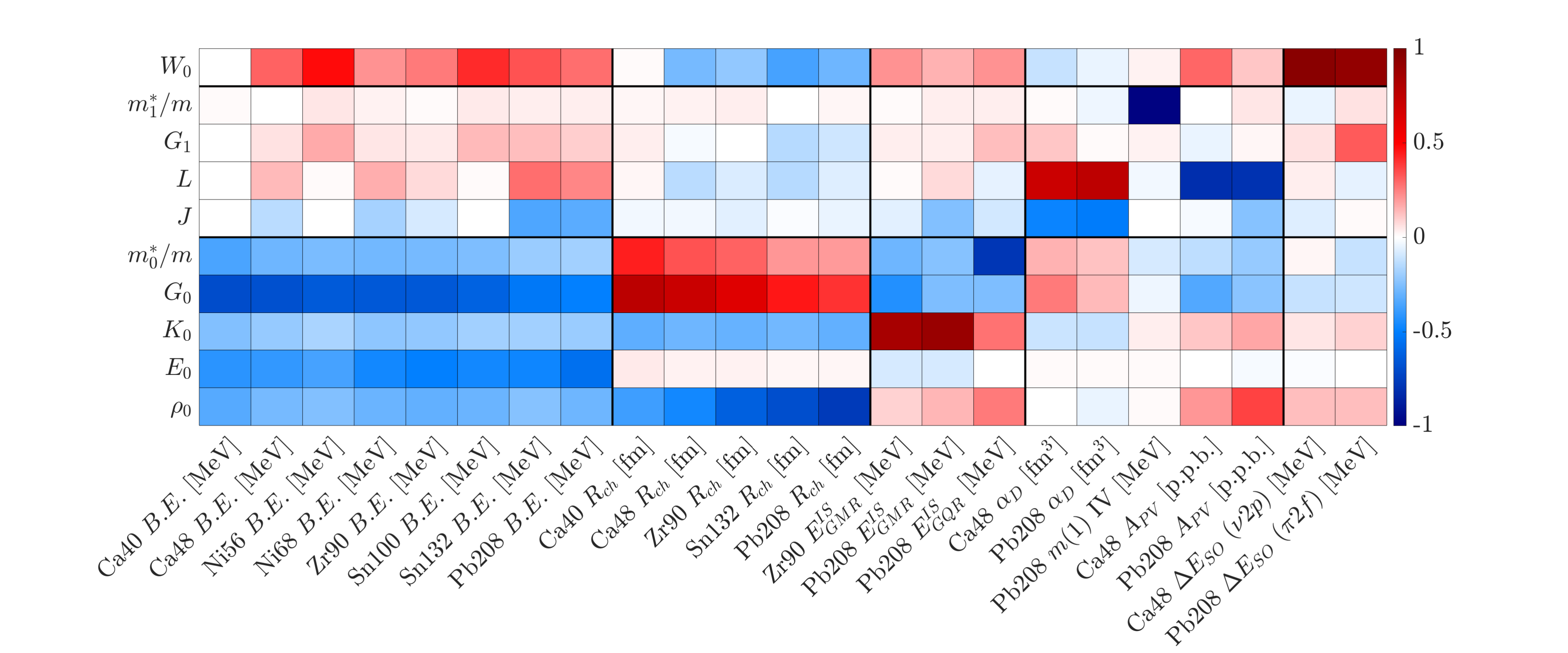}
  \caption{Correlations between parameters and observables. 
           The two thick horizontal lines divide the spin-orbit,isoscalar and isovector parameters, while the thick vertical lines separate  the different  groups of observables  included in our analysis according to their type .}
  \label{fig:correlations}
\end{figure*}

Finally, we turn to the correlations between the model parameters and the observables.
To study those, we take the model results over the training grid, which spans the whole parameter space, and compute the Pearson correlation factors of all possible combination observable-parameter; 
we thus find a correlation matrix, which we show in Fig.~\ref{fig:correlations}.
The observables are along the columns, grouped by their type and ordered from the lightest to the heaviest nucleus, while the parameters are on the rows, divided into isoscalar, isovector, and spin-orbit. 
The value of the correlation coefficient is in a colored scale, going from dark blue (-1, total anti-correlation) to dark red (1, total correlation).
0, no correlation, is mapped in white.
 
The energy at saturation $E_0$ is mainly anti-correlated with the binding energies, and very little with the charge radii.
On the other hand, the saturation density $\rho_0$ is anti-correlated with the charge radii (and especially with that of $^{208}$Pb). 
These correlations can be expected on quite general grounds. If one increases $E_0$, nuclei are overbound and the ``anti-''correlation is merely a result of the sign convention on the binding energy; at the same time, a higher (lower) saturation density leads to more compact (more dilute) nuclei.
The compressibility $K_0$ is highly correlated with both $E_{\rm GMR}^{\rm IS}$'s, as is well known and has been already discussed.
The isoscalar surface parameter $G_0$ is correlated well with the ground state observables (slightly more to those of lighter nuclei, where the surface plays a stronger role), while the isoscalar effective mass $m^*_0/m$ is mainly constrained by $E_{\rm GQR}^{\rm IS}$ as expected.

As for the isovector parameters, $J$ is slightly anti-correlated with the polarizabilities $\alpha_D$; 
on the other hand, $L$ is strongly correlated with $\alpha_D$ \cite{Piekarewicz2012, Roca-Maza2013_2, Roca-Maza2015} and anti-correlated with $A_{PV}$ \cite{Roca-Maza2011-2}.
These correlations have of course a secondary effect on the distribution of $J$. 
For the isovector surface parameter $G_1$, no single observable acts as a stringent constraint, which explains its rather wide distribution.
The situation is the opposite for the isovector effective mass $m^*_1/m$, which is specifically affected by the EWSR $m(1)$ of the $^{208}$Pb IVGDR.
Finally, the spin-orbit parameter $W_0$ is highly correlated to the spin-orbit splittings, but with the binding energies as well.

\section{Conclusions} 
\label{sec:conclusions}
In this study, we have investigated the traditional Skyrme ansatz within the Bayesian inference framework, using several properties of nuclei, both for the ground state and excited states (isoscalar and isovector collective resonances), as physical constraints.
In particular, we have included the parity-violating asymmetry and the dipole polarizability, both measured for $^{48}$Ca and $^{208}$Pb, which are generally considered to be in tension with each other. 

We have carried out a sequence of inferences, by gradually including the different types of constraints. 
This allows us to pin down more clearly the role played by the various observables in shaping the posterior distributions.
The final result is a ten-dimensional probability distribution of the Skyrme parameters, expressed in terms of nuclear matter parameters and the EDF surface and spin-orbit parameters.  
The marginalized posterior distributions for each parameter are similar to previous results in the literature, except for $L$, for which we found lower values.
The combined effect of the constraints from masses, radii, and especially dipole polarizabilities and $A_{PV}$($^{48}$Ca), lies at the basis of this result, and it is not counterbalanced by the $A_{PV}$($^{208}$Pb).

The posterior distributions of observables are compatible with the experimental values. 
The only one showing a significant deviation is the parity-violating asymmetry of $^{208}$Pb.
This is not unexpected, since the $L$ distribution is shaped mainly by $\alpha_D$, and high $L$ values are needed to meet the result of the PREX-II experiment.

This work can be the starting point for further analysis. 
On the one hand, we envision studying if the present distributions for the EoS parameters can be reconciled with those extracted from the observables associated with a broader range of densities, like those coming from neutron stars. As this may not happen, generalizations of the Skyrme ansatz should be considered. However, the present work has several possible improvements even when sticking only to finite nuclei, or ordinary nuclear densities. As discussed throughout the text, one should probably include more observables in the inference. It should be checked whether open-shell, deformed nuclei will give a different bias (this will call for serious consideration of the pairing channel). More excited states may also be considered. Last but not least, the time-odd part of the EDF has not been addressed in this work.

\begin{acknowledgments}
XRM acknowledges support by MICIU/AEI/10.13039/501100011033 and by FEDER UE
through grants PID2023-147112NB-C22; and through the ``Unit of Excellence Mar\'ia de Maeztu 2020-2023" award to the Institute of Cosmos Sciences, grant CEX2019-000918-M. Additional support is provided by the Generalitat de Catalunya (AGAUR) through grant 2021SGR01095. 
\end{acknowledgments}

\bibliography{bibliography}

\appendix

\section{Validation}
\label{app:validation}
We go through our validation process, using the final inference ``$+A_{PV}$ ($^{208}$Pb and $^{48}$Ca)'' including all the observables as an example.
As mentioned in the main text, to assess the emulator's performance, we extract 250 samples from the posterior distribution of parameters and compare the model results with those obtained by using the emulator.
We require that the difference between the model and the emulator results should be smaller than the error we used for the inferences for at least 95\% of the points (Tab.~\ref{tab:obs}).

\begin{figure*}
  \includegraphics[width=\linewidth]{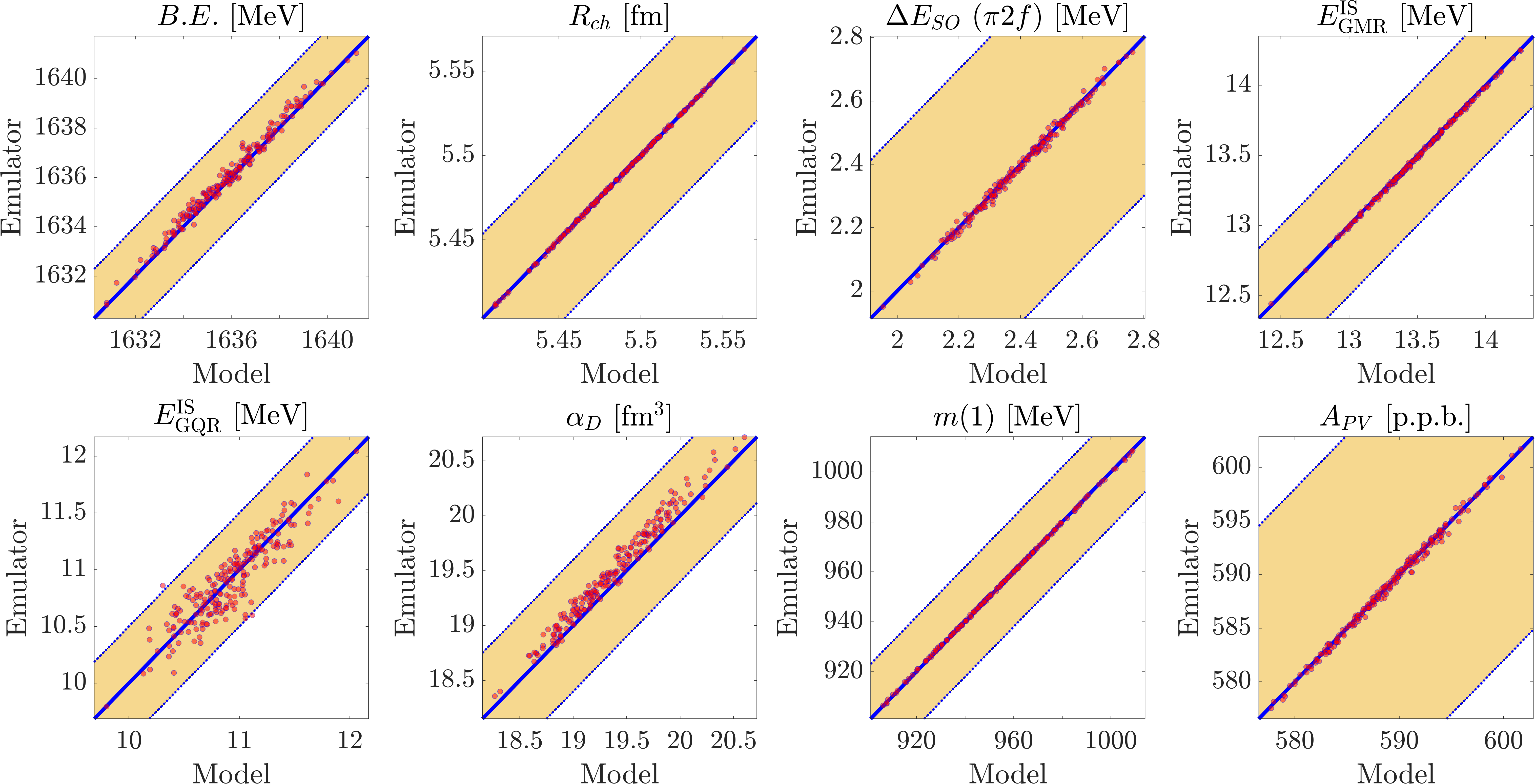}
  \caption{Validation plot for the $^{208}$Pb observables. \label{fig:validation_plot}}
\end{figure*}

\begin{table}
	\centering
	\caption{Emulator performance (``$+A_{PV}$($^{208}$Pb and $^{48}$Ca)'' inference).}
	\label{tab:full_run_new_validation}
	\begin{tabular}{cccc}
	\hline
	\multicolumn{4}{c}{Ground-state properties} \\ 
	\hline
	   & $B.E.$ & $R_{\rm ch}$ & $\Delta E_{\rm SO}$   \\ 
	\hline
	$^{208}$Pb &        0 \% &        0 \% &        0 \%  \\ 
	$^{48}$Ca &        0 \% &        0 \% &        0 \%  \\ 
	$^{40}$Ca &        0 \% &        0 \% & - \\ 
	$^{56}$Ni &        0 \% & -& -\\ 
	$^{68}$Ni &        0 \% & -& -\\ 
	$^{100}$Sn &        0 \% & -& -\\ 
	$^{132}$Sn &        0 \% &        0 \%& - \\ 
	$^{90}$Zr &        0 \% &        0 \%& -\\ 
	\end{tabular}\\ 
	\vspace{0.3cm}
	\begin{tabular}{ccc}
	\hline
	\multicolumn{3}{c}{Isoscalar resonances} \\ 
	\hline
	   & $E_{\rm GMR}^{\rm IS}$  & $E_{\rm GQR}^{\rm IS}$  \\ 
	\hline
	$^{208}$Pb &        0 \% &      1.0 \%  \\ 
	$^{90}$Zr &        0 \% & -\\ 
	\end{tabular}\\ 
	\vspace{0.3cm}
	\begin{tabular}{cccc}
	\hline
	\multicolumn{4}{c}{Isovector properties} \\ 
	\hline
	   & $\alpha_D$ & $m(1)$ & $A_{PV}$  \\ 
	\hline
	$^{208}$Pb &        0 \% &        0 \% &        0 \%  \\ 
	$^{48}$Ca &        0 \% & -&        0 \%    \\ 
	\hline
	\end{tabular}\\ 
\end{table}

In Fig.~\ref{fig:validation_plot}, we show a graphic representation of our analysis for the observables of $^{208}$Pb. 
The x-coordinate of each red point corresponds to the model result, while the y-coordinate to its emulator's counterpart. 
The blue line is $x = y$, while the two dotted blue lines that delimit the shaded yellow region are $ x = y \pm \Delta$, where $\Delta$ are the errors associated with each observable (Tab.~\ref{tab:obs}).
If at least 95\% of the points fall between the two lines, the emulator satisfies our discrepancy requirement.
As we can see, there is no point outside the accepted boundaries 
except in the case of $E_{\rm GQR}^{\rm IS}$, where only 2 points out of 250 lie outside the allowed region.

In Tab.~\ref{tab:full_run_new_validation}, we report the discrepancy percentages of all the observables for the ``$+A_{PV}$ ($^{208}$Pb and $^{48}$Ca)'' inference.
As we can see, the emulator always meets the validation criterion, and the
only case where the discrepancy is greater than 0 is just for $E_{\rm GQR}^{\rm IS}$.
For the other inferences, the validation is likewise satisfactory.
In the following Tables (\ref{tab:gs_only_validation}, \ref{tab:gs_so_only_validation_new},\ref{tab:gs_so_plus_alfaD_new_validation}, \ref{tab:no_apv_validation_new}, \ref{tab:no_apv_validation_ca_new} and \ref{tab:no_apv_validation_pb_new}) there is the data. 
All observables in all inferences have a discrepancy of 0\%.
The only exception is the already mentioned $E_{\rm GQR}^{\rm IS}$, which varies between 0.4\% and 4.8\%.

\begin{table}
	\centering
	\caption{Emulator performance (``$B.E.\; R_{ch}$'' inference).}
	\label{tab:gs_only_validation}
	\begin{tabular}{ccc}
	\hline
	\multicolumn{3}{c}{Ground-state properties} \\ 
	\hline
	   & $B.E.$ & $R_{\rm ch}$   \\ 
	\hline
	$^{208}$Pb &        0 \% &        0 \%    \\ 
	$^{48}$Ca &        0 \% &        0 \%   \\ 
	$^{40}$Ca &        0 \% &        0 \%  \\ 
	$^{56}$Ni &        0 \% & -\\ 
	$^{68}$Ni &        0 \% & -\\ 
	$^{100}$Sn &        0 \% & -\\ 
	$^{132}$Sn &        0 \% &        0 \% \\ 
	$^{90}$Zr &        0 \% &        0 \% \\ 
	\end{tabular}\\ 
\end{table}

\begin{table}
	\centering
	\caption{Emulator performance (``$+\Delta E_{SO}$'' inference).}
	\label{tab:gs_so_only_validation_new}
	\begin{tabular}{cccc}
	\hline
	\multicolumn{4}{c}{Ground-state properties} \\ 
	\hline
	   & $B.E.$ & $R_{\rm ch}$ & $\Delta E_{\rm SO}$   \\ 
	\hline
	$^{208}$Pb &        0 \% &        0 \% &        0 \%  \\ 
	$^{48}$Ca &        0 \% &        0 \% &        0 \%  \\ 
	$^{40}$Ca &        0 \% &        0 \% & -\\ 
	$^{56}$Ni &        0 \% & -& -\\ 
	$^{68}$Ni &        0 \% & -& -\\ 
	$^{100}$Sn &        0 \% & -& -\\ 
	$^{132}$Sn &        0 \% &        0 \%& - \\ 
	$^{90}$Zr &        0 \% &        0 \%& -\\ 
	\end{tabular}\\ 
\end{table}

\begin{table}
	\centering
	\caption{Emulator performance (``$+\alpha_D$'' inference).}
	\label{tab:gs_so_plus_alfaD_new_validation}
	\begin{tabular}{cccc}
	\hline
	\multicolumn{4}{c}{Ground-state properties} \\ 
	\hline
	   & $B.E.$ & $R_{\rm ch}$ & $\Delta E_{\rm SO}$  \\ 
	\hline
	$^{208}$Pb &        0 \% &        0 \% &        0 \%  \\ 
	$^{48}$Ca &        0 \% &        0 \% &        0 \%  \\ 
	$^{40}$Ca &        0 \% &        0 \% & -\\ 
	$^{56}$Ni &        0 \% & -& -\\ 
	$^{68}$Ni &        0 \% & -& -\\ 
	$^{100}$Sn &        0 \% & -& -\\ 
	$^{132}$Sn &        0 \% &        0 \%& - \\ 
	$^{90}$Zr &        0 \% &        0 \%& -\\ 
	\end{tabular}\\ 
	\vspace{0.3cm}
	\begin{tabular}{cccc}
	\hline
	\multicolumn{4}{c}{Isovector properties} \\ 
	\hline
	   & $\alpha_D$ & $m(1)$ & $A_{PV}$ \\ 
	\hline
	$^{208}$Pb &        0 \% & \color{Red}{X} & \color{Red}{X}  \\ 
	$^{48}$Ca &        0 \% & -& \color{Red}{X}   \\ 
	\hline
	\end{tabular}\\ 
\end{table}

\begin{table}
	\centering
	\caption{Emulator performance (``$+GR$'' inference.)}
	\label{tab:no_apv_validation_new}
	\begin{tabular}{cccc}
	\hline
	\multicolumn{4}{c}{Ground-state properties} \\ 
	\hline
	   & $B.E.$ & $R_{\rm ch}$ & $\Delta E_{\rm SO}$  \\ 
	\hline
	$^{208}$Pb &        0 \% &        0 \% &        0 \% \\ 
	$^{48}$Ca &        0 \% &        0 \% &        0 \% \\ 
	$^{40}$Ca &        0 \% &        0 \% & -\\ 
	$^{56}$Ni &        0 \% & -& -\\ 
	$^{68}$Ni &        0 \% & -& -\\ 
	$^{100}$Sn &        0 \% & -& -\\ 
	$^{132}$Sn &        0 \% &        0 \%& - \\ 
	$^{90}$Zr &        0 \% &        0 \%& -\\ 
	\end{tabular}\\ 
	\vspace{0.3cm}
	\begin{tabular}{ccc}
	\hline
	\multicolumn{3}{c}{Isoscalar resonances} \\ 
	\hline
	   & $E_{\rm GMR}^{\rm IS}$  & $E_{\rm GQR}^{\rm IS}$   \\ 
	\hline
	$^{208}$Pb &        0 \% &      4.8 \%  \\ 
	$^{90}$Zr &        0 \% & -\\ 
	\end{tabular}\\ 
	\vspace{0.3cm}
	\begin{tabular}{cccc}
	\hline
	\multicolumn{4}{c}{Isovector properties} \\ 
	\hline
	   & $\alpha_D$ & $m(1)$ & $A_{PV}$  \\ 
	\hline
	$^{208}$Pb &        0 \% &        0 \% & \color{Red}{X}  \\ 
	$^{48}$Ca &        0 \% & -& \color{Red}{X}    \\ 
	\hline
	\end{tabular}\\ 
\end{table}

\begin{table}
	\centering
	\caption{Emulator performance (``$+A_{PV}$ ($^{208}$Pb only)'' inference).}
	\label{tab:no_apv_validation_ca_new}
	\begin{tabular}{cccc}
	\hline
	\multicolumn{4}{c}{Ground-state properties} \\ 
	\hline
	   & $B.E.$ & $R_{\rm ch}$ & $\Delta E_{\rm SO}$   \\ 
	\hline
	$^{208}$Pb &        0 \% &        0 \% &        0 \%  \\ 
	$^{48}$Ca &        0 \% &        0 \% &        0 \%  \\ 
	$^{40}$Ca &        0 \% &        0 \% & - \\ 
	$^{56}$Ni &        0 \% & -& -\\ 
	$^{68}$Ni &        0 \% & -& -\\ 
	$^{100}$Sn &        0 \% & -& -\\ 
	$^{132}$Sn &        0 \% &        0 \%& - \\ 
	$^{90}$Zr &        0 \% &        0 \%& -\\ 
	\end{tabular}\\ 
	\vspace{0.3cm}
	\begin{tabular}{ccc}
	\hline
	\multicolumn{3}{c}{Isoscalar resonances} \\ 
	\hline
	   & $E_{\rm GMR}^{\rm IS}$  & $E_{\rm GQR}^{\rm IS}$  \\ 
	\hline
	$^{208}$Pb &        0 \% &      0.4 \%  \\ 
	$^{90}$Zr &        0 \% & -\\ 
	\end{tabular}\\ 
	\vspace{0.3cm}
	\begin{tabular}{cccc}
	\hline
	\multicolumn{4}{c}{Isovector properties} \\ 
	\hline
	   & $\alpha_D$ & $m(1)$ & $A_{PV}$   \\ 
	\hline
	$^{208}$Pb &        0 \% &        0 \% &        0 \%  \\ 
	$^{48}$Ca &        0 \% & -& \color{Red}{X}    \\ 
	\hline
	\end{tabular}\\ 
\end{table}

\begin{table}
	\centering
	\caption{Emulator performance (``$+A_{PV}$ ($^{48}$Ca only)'' inference).}
	\label{tab:no_apv_validation_pb_new}
	\begin{tabular}{cccc}
	\hline
	\multicolumn{4}{c}{Ground-state properties} \\ 
	\hline
	   & $B.E.$ & $R_{\rm ch}$ & $\Delta E_{\rm SO}$   \\ 
	\hline
	$^{208}$Pb &        0 \% &        0 \% &        0 \%  \\ 
	$^{48}$Ca &        0 \% &        0 \% &        0 \%  \\ 
	$^{40}$Ca &        0 \% &        0 \% & - \\ 
	$^{56}$Ni &        0 \% & -& -\\ 
	$^{68}$Ni &        0 \% & -& -\\ 
	$^{100}$Sn &        0 \% & -& -\\ 
	$^{132}$Sn &        0 \% &        0 \%& - \\ 
	$^{90}$Zr &        0 \% &        0 \%& -\\ 
	\end{tabular}\\ 
	\vspace{0.3cm}
	\begin{tabular}{ccc}
	\hline
	\multicolumn{3}{c}{Isoscalar resonances} \\ 
	\hline
	   & $E_{\rm GMR}^{\rm IS}$  & $E_{\rm GQR}^{\rm IS}$  \\ 
	\hline
	$^{208}$Pb &        0 \% &      0.8 \%  \\ 
	$^{90}$Zr &        0 \% & - \\ 
	\end{tabular}\\ 
	\vspace{0.3cm}
	\begin{tabular}{cccc}
	\hline
	\multicolumn{4}{c}{Isovector properties} \\ 
	\hline
	   & $\alpha_D$ & $m(1)$ & $A_{PV}$   \\ 
	\hline
	$^{208}$Pb &        0 \% &        0 \% & \color{Red}{X} \\ 
	$^{48}$Ca &        0 \% & -&        0 \%   \\ 
	\hline
	\end{tabular}\\ 
\end{table}

\end{document}